\documentclass[openacc]{rstransa}
\usepackage{subfig}
\usepackage{hyperref}

\titlehead{Research}

\begin{document}

\title{Radiative hydrodynamic simulations of FIP fractionation in solar flares}

\author{
Jeffrey W. Reep$^{1}$, Luke Fushimi Benavitz$^{1}$, Andy S.H. To$^{2}$,
David H. Brooks$^{3,4}$, J. Martin Laming$^{5}$, Patrick Antolin$^{6}$,
David M. Long$^{7,8}$, Deborah Baker$^{4}$
}

\address{
$^{1}$Institute for Astronomy, University of Hawai’i, USA\\
$^{2}$ESTEC, European Space Agency, The Netherlands\\
$^{3}$Computational Physics, Inc., USA\\
$^{4}$University College London, Mullard Space Science
Laboratory, UK\\
$^{5}$Space Science Division, Naval Research Laboratory, USA\\
$^{6}$Northumbria University, Newcastle upon Tyne, UK\\
$^{7}$Dublin City University, Glasnevin Campus, Dublin, Ireland\\
$^{8}$Dublin Institute for Advanced Studies, Dublin, Ireland\\
}

\subject{astrophysics, stars, spectroscopy}


\corres{Jeffrey Reep\\
\email{reep@hawaii.edu} }

\begin{abstract}
Elemental abundances in solar flares are observed to vary both spatially and temporally, but the underlying mechanisms remain poorly understood.   There is an interplay between advection and the preferential acceleration of low first ionization potential (FIP) elements that likely shapes the observed abundance distributions.  Models of the FIP effect predict enhancements near loop footpoints that diffuse upward over time.  We simulate strong evaporation events that advect this low-FIP enhancement into the corona.  When the enhancement is sharply peaked, the corona does not become fractionated, exhibiting only a localized abundance peak near the loop apex that facilitates coronal rain formation.  In contrast, a broad enhancement with relatively weak heating yields a uniformly fractionated corona, which is not sufficient for coronal rain formation.  As the heating rate increases, the low-FIP enhanced plasma is increasingly compressed toward the loop apex, and rain is able to form.  These results suggest a potential observational correlation between the presence and amount of coronal rain, the strength of flare heating, and the fractionation process itself.
\end{abstract}

\maketitle


\section{Introduction}

In the quiescent solar corona, elements with low first ionization potential (FIP) $\lesssim 10$ eV are enhanced in abundance relative to their photospheric values by factors of $\approx 3$–5, a phenomenon known as the FIP effect \cite{laming2015}.  This enhancement has been observed in closed coronal loops \cite{widing2001,delzanna2014} and the solar wind \cite{bochsler2007,brooks2012}.  For many years, observations of solar flares typically found photospheric abundances \cite{feldman1990,sylwester2010,warren2014}, but more recent studies have revealed more complex and variable behavior that appears to differ from event to event.  Elemental abundances in flares have been observed to vary both spatially (\textit{e.g.} \cite{doschek2015,doschek2018,baker2019,to2021,to2024}) and temporally (\textit{e.g.} \cite{sylwester1998,katsuda2020,mondal2021,nama2023,sylwester2023,woods2023}).  

Most (magneto)hydrodynamic simulations of the solar atmosphere assume fixed elemental abundances, typically coronal for active region (AR) modeling and photospheric for flare modeling.  However, it has long been recognized that assumed abundances significantly influence the radiative loss rate \cite{cox1969,cook1989,landi1999}, and therefore the cooling rate of the plasma.  As a result, the cooling rate may act indirectly as a diagnostic of the FIP and inverse FIP effects \cite{brooks2018,brooks2025}.  Recent work has begun to explore the limitations of assuming fixed abundances, which have a major impact on the dynamics.

Time-variable abundances were recently implemented \cite{reep2024} in the \texttt{ebtel++} code, a zero-dimensional, coronally-averaged hydrodynamic model of closed coronal loops \cite{barnes2016,barnes2025}.  The model assumed that loops initially have coronal abundances, and chromospheric evaporation carries photospheric material upwards into the corona, thereby modifying the loop's emissivity over time.  Strong heating events induce strong evaporation, reducing the coronal abundances to photospheric values, while weaker events yield intermediate abundances.  Since the \texttt{ebtel++} code does not have spatial resolution, the ability to make observational comparisons is limited.

To improve upon this, spatiotemporally-variable abundances were then implemented \cite{benavitz2025} in a field-aligned model of the solar atmosphere, \texttt{HYDRAD}, which solves the radiative hydrodynamic equations for a plasma constrained to a magnetic flux tube, for either closed loops or the solar wind \cite{bradshaw2003,bradshaw2013,scott2022}\footnote{https://github.com/rice-solar-physics/HYDRAD}.  The simulations similarly assumed initially that there were coronal abundances in the corona and photospheric abundances in the chromosphere, such that evaporation would fill the loop with photospheric material.  In impulsive heating events (both nanoflare and electron beam heating), evaporation pushes the initial coronal plasma towards the apex, producing a localized enhancement in low-FIP elements.  Because of this enhancement, the local radiation is stronger than in the surrounding plasma, leading to faster cooling locally, directly causing the formation of a coronal condensation \cite{benavitz2025}, which is otherwise absent from impulsively-heated loop simulations \cite{reep2020}, although cases with magnetic dips have also been found to form condensations \cite{huang2021,yoshihisa2025}.  This enhancement of radiation is also consistent with observations of looptop bright points commonly seen in low-FIP element spectral line emission in flares \cite{doschek1996,doschek2018,to2024}.

Coronal rain forms as a result of runaway cooling caused by a rapid increase in the radiative loss rate.  Rain is commonly seen in images at chromospheric or transition region temperatures (\textit{e.g.} H$\alpha$ \cite{scullion2016,jing2016}, He I \cite{schad2018}, Si IV \cite{kohutova2016}, Ca II \cite{antolin2012}), located in the corona, as blobs of material that precipitate down field lines at speeds of $\approx 50 - 150$ km s$^{-1}$ \cite{antolin2012,kohutova2016,schad2018,antolin2023}.  Thought to form by thermal instability \cite{field1965,xia2011}, the onset of the process becomes detectable around $10^{6}$ K \cite{scullion2016,brooks2024}, rapidly cooling to the chromospheric temperatures characteristic of rain, below $10^{4}$ K \cite{antolin2012,antolin2015,scullion2016}.  The densities are typically of order $10^{10}$ to $10^{12}$ cm$^{-3}$ \cite{antolin2015,scullion2016}.  Additionally, spectral lines are significantly broadened in rain as compared to the loops themselves, suggesting that the formation process may be turbulent \cite{brooks2024}.

It has long been known that thermal non-equilibrium (TNE) can produce coronal rain \cite{antiochos1991,muller2003}, where quasi-steady heating near the footpoints of loops \cite{froment2018} or open field lines \cite{scott2024} drives a slow accumulation of mass higher in the corona, and a localized imbalance between heating and radiative losses then initiates the cooling that leads to rain.  Although this process can cause rain, it requires relatively long heating timescales of perhaps 30 or more minutes.  By contrast, rain can form with short heating timescales (seconds) when driven by abundance variability.  A recent flare observation found a possible difference between the abundances in a post-flare loop and the rain that formed there \cite{brooks2024}, suggesting that variability in low-FIP elemental abundances may contribute to the formation of coronal rain.

It is therefore important to determine (1) how abundances change during flares, and (2) how these changes relate to the formation of coronal rain.  Magnetic reconnection events that heat the plasma also generate Alfv\'enic waves that can drive the fractionation that produces coronal abundances \cite{laming2009,laming2012}. Ponderomotive acceleration in these models predicts a sharp enhancement in low-FIP elements near loop footpoints, suggesting that evaporation could carry fractionated material into the corona \cite{laming2021}.  Since the ponderomotive force acts only on ions, and ionization fractions vary with chromospheric depth, the depth from which plasma evaporates could influence the resultant abundances in the corona.  In this work, we explore this idea directly: how does the fractionation depth in the chromosphere affect the evolution of abundances in the corona?  We assume the presence of low-FIP elemental abundance spikes near the footpoints of loops and examine how evaporation transports this material into the corona.

\section{Transport of low-FIP elements}

We consider here the advection of low-FIP elements (FIP $< 10$\ eV), using the model described by \cite{benavitz2025}\footnote{https://github.com/jwreep/HYDRAD/tree/Abundances}.  First, we define the abundance factor $f$ as the local deviation from photospheric abundances $\Omega_{\text{photo}}$:
\begin{equation}
    f(s,t) = \frac{\Omega(s,t)}{\Omega_{\text{photo}}}
\end{equation}
\noindent We assume that the abundance factor $f$ is the same for all low-FIP elements in this model, though this is an oversimplification that should be generalized in the future.  We do not vary the abundances of elements with FIP $> 10$\ eV, assuming that they are photospheric in value everywhere.  

The continuity equation for low-FIP elements can then be written \cite{benavitz2025}:
\begin{equation}
\partial_{t} (\rho f) = -\frac{1}{A} \partial_{s}(A\rho f v)
\end{equation}
\noindent where $\rho$ is the mass density, $A$ the cross-sectional area, and $v$ the bulk flow velocity.  This equation advects the low-FIP elements, but cannot cause fractionation itself.  

\texttt{HYDRAD} \cite{bradshaw2003,bradshaw2013} solves the conservation of mass, momentum, and energy for plasma constrained to flow along a magnetic flux tube.  Radiative losses in the corona are treated by calculating the total emissivity, summing over all ions, using the \texttt{CHIANTI} atomic database \cite{dere1997}, version 10.1 \cite{dere2023}.  We modify the local abundances of low-FIP elements by $f(s,t)$, which affects the radiative loss rate in each grid cell.    We assume that $f = 1$ corresponds to photospheric, using the values measured by \cite{asplund2009}.  We include the 15 most abundant elements in the radiation calculation.

\section{Simulations}

We examine \texttt{HYDRAD} simulations of coronal loops with total length $2L = 50$\ Mm subjected to heating by an electron beam injected on both legs of the loop.  The loops are initially in hydrostatic equilibrium, with the coronal temperatures and densities determined by solving the hydrostatic equations, integrated from the base of the transition region.  The chromospheric temperature profile follows the VAL C model \cite{vernazza1981}, with an approximation to optically-thick cooling rates \cite{carlsson2012}.  The heating rate due to an electron beam is determined by the distribution of injected electrons, the column density, and the ionization fraction \cite{emslie1978,hawley1994}.  We assume an electron distribution characteristic of a moderate flare \cite{holman2011,hannah2011}, with a low-energy cutoff $E_{c} = 15$\ keV, spectral index $\delta = 5$, for 10 s of constant energy fluxes $F = [2 \times 10^{10}, 5 \times 10^{10}, 10^{11}, 2 \times 10^{11}]$\ erg s$^{-1}$ cm$^{-2}$.   

The simulations in \cite{reep2024,benavitz2025} assumed that the loops had already been fractionated, that is, that the coronal portion of the loops had coronal abundances with $f = 4$.  This is an assumption, however, that has not been observationally validated in flare loops.  In this work, we examine two scenarios proposed for how elemental abundances might fractionate and vary in flares.  The results of \cite{laming2012} and \cite{laming2019} (as well as the Laming and To et al. papers in this issue) suggest that the fractionation of each species varies with height, depending strongly on the chromospheric density and temperature profiles as well as the wave spectra present.  The cases with narrow spikes are similar to the predictions of \cite{laming2012} where the fractionation is primarily driven by resonant Alfv\'enic waves propagating downwards from the corona, while the cases with wide spikes are similar to the predictions with non-resonant waves. 

In our simulations, we introduce spikes in the abundance factor $f$ centered 2 Mm above the photosphere at each footpoint (upper chromosphere), with amplitude $f = 5$, and assume photospheric abundances ($f = 1$) elsewhere.\footnote{The initial amplitude of the spike is less critical than the width, which sets how much total low-FIP enhanced plasma can be moved into the corona.  We have included two supplementary simulations with $f = 4$ in the Zenodo cache that show that the basic results still hold.  Of course, if the amplitude were too small (say $f=2$ or less), then the radiation may not be strong enough to produce rain.}  We consider two cases with different Gaussian widths: (1) 100 km width, sharply spiked as predicted by models of the ponderomotive acceleration \cite{laming2021}, and (2) 500 km width, wide spikes with enhancement of low-FIP enhanced plasma extending much deeper into the chromosphere (a scenario suggested by \cite{brooks2024} and consistent with the calculations of \cite{laming2025} in these proceedings).  

The first case, with narrow spikes, is shown in Figure \ref{fig:narrow}.  Each set of 5 panels shows one simulation.  The panel at left shows the abundance factor $f(s,t)$ for the first 300 s of the simulation, at a 10-second cadence, from blue to green to yellow in time.  The panels on the right show time-distance plots of the electron temperature, electron density, bolometric radiative loss rate, and bulk flow velocity.  In the velocity plots, blue means motion towards the apex, red away from the apex.  The colorbars on each plot indicate the magnitudes of each respective quantity.  The heating begins at the start of each simulation, quickly heating the upper chromosphere, raising the temperature and pressure there, inducing evaporative flows into the corona.  These flows carry material and energy into the loop, raising the coronal temperature and density.  As a result of the sharp increase in density, the radiative loss rate increases across the entirety of the corona.  Initially cooling primarily through conduction, as the temperature falls, the cooling becomes driven predominantly by radiation and flows out of the loop \cite{cargill1995,bradshaw2005,bradshaw2010}.  When the temperature falls to just below 1 MK, the loop catastrophically collapses, with the temperature falling rapidly to chromospheric values and the plasma evacuating the loop rapidly \cite{cargill2013}.
\begin{figure}[!h]
\centering
\includegraphics[width=\textwidth]{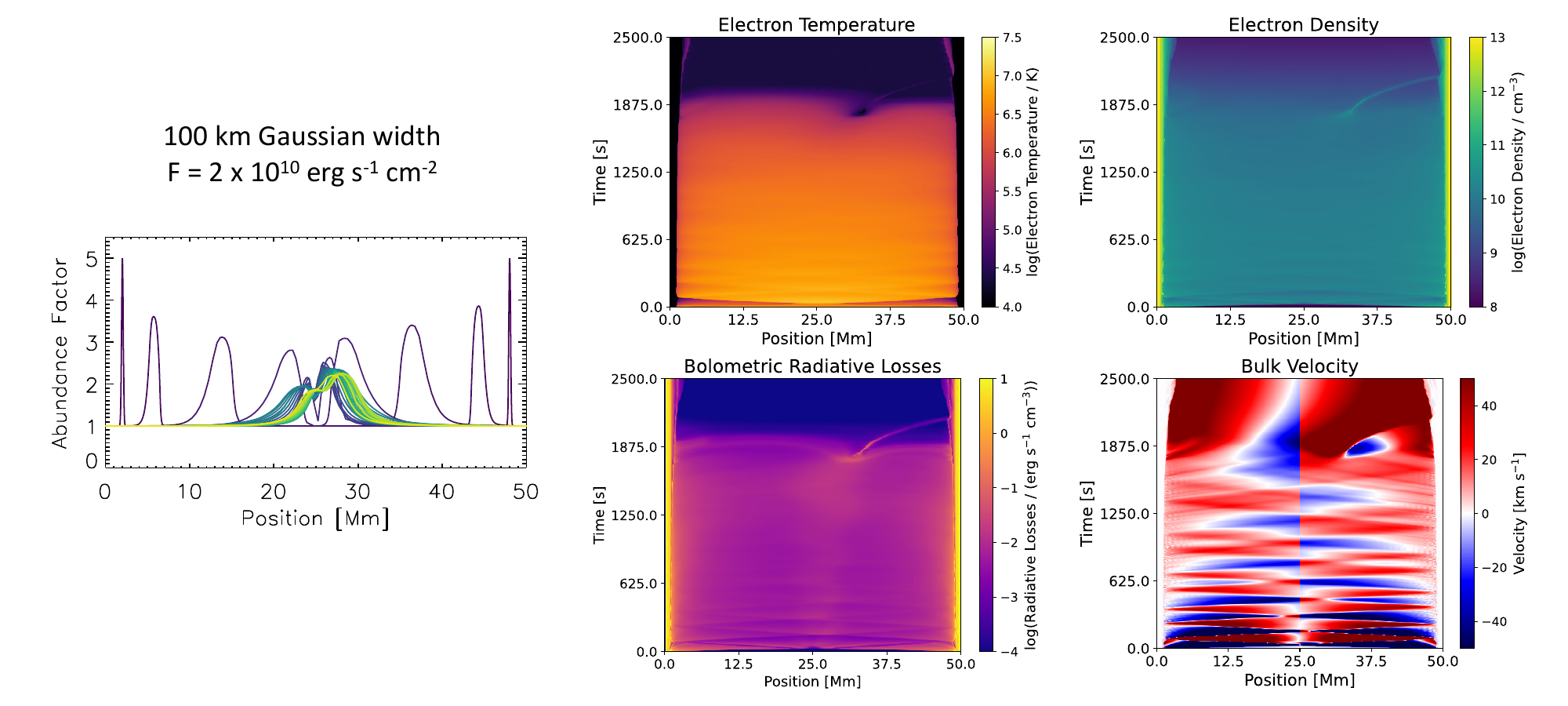}
\includegraphics[width=\textwidth]{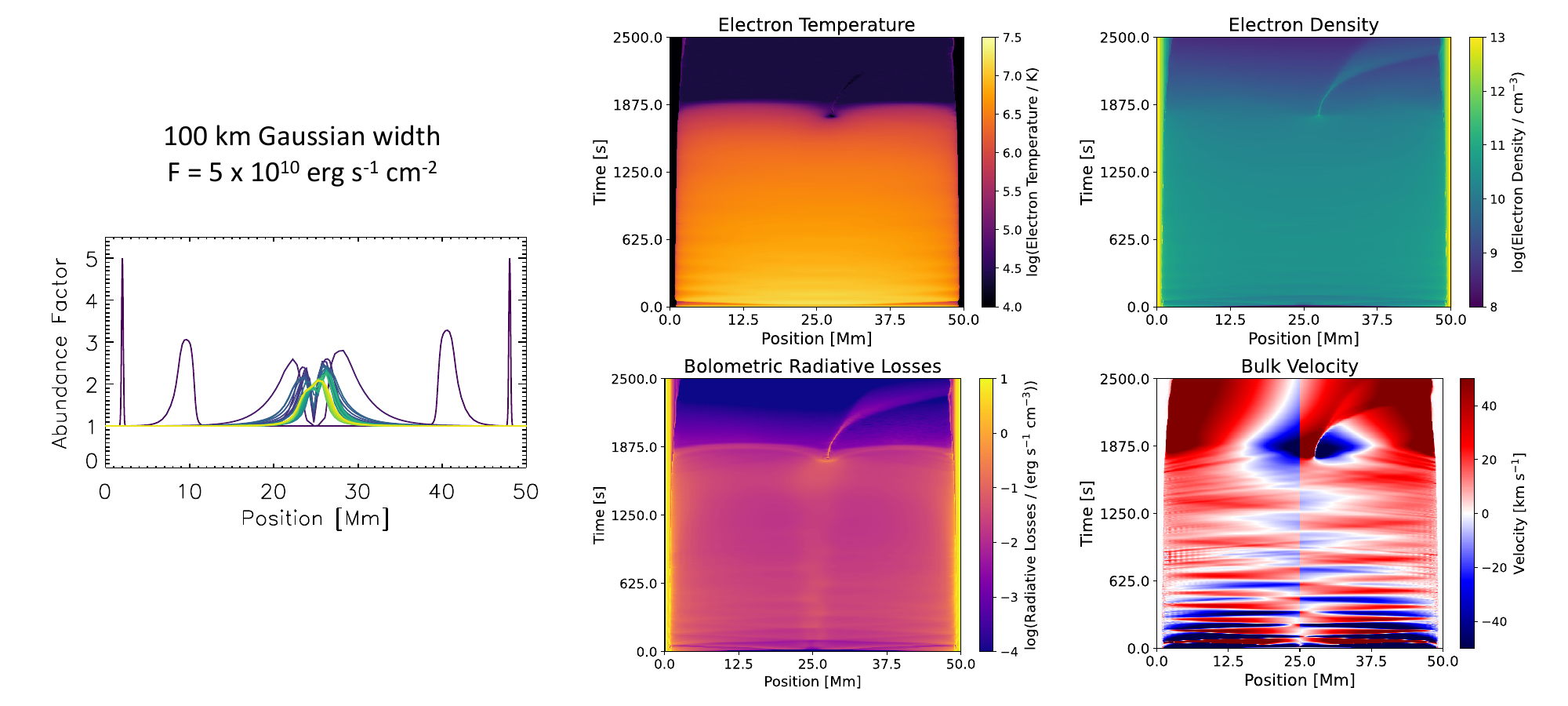}
\caption{Simulations with narrow spikes in the abundance factor $f$ at the footpoints of the loops.  Each set of five plots shows one simulation.  The panel at left in each case shows the abundance factor, from times 0 to 300 s in the simulation at a 10-second cadence (blue to green to yellow in time).  The panels on the right show time-distance plots of the electron temperature, electron density, bolometric radiative loss rate, and the bulk flow velocity (where blue means towards the apex, red away from the apex), tracking the evolution of the loop in each simulation.  In all cases, a spike in $f$ forms near the apex, enhancing the radiative loss rate there and causing the formation of a coronal rain event (seen as high density, low temperature regions late in the simulations).  (Figure continued on next page.)}
\label{fig:narrow}
\end{figure}
\begin{figure}\ContinuedFloat
    \centering
    \includegraphics[width=\textwidth]{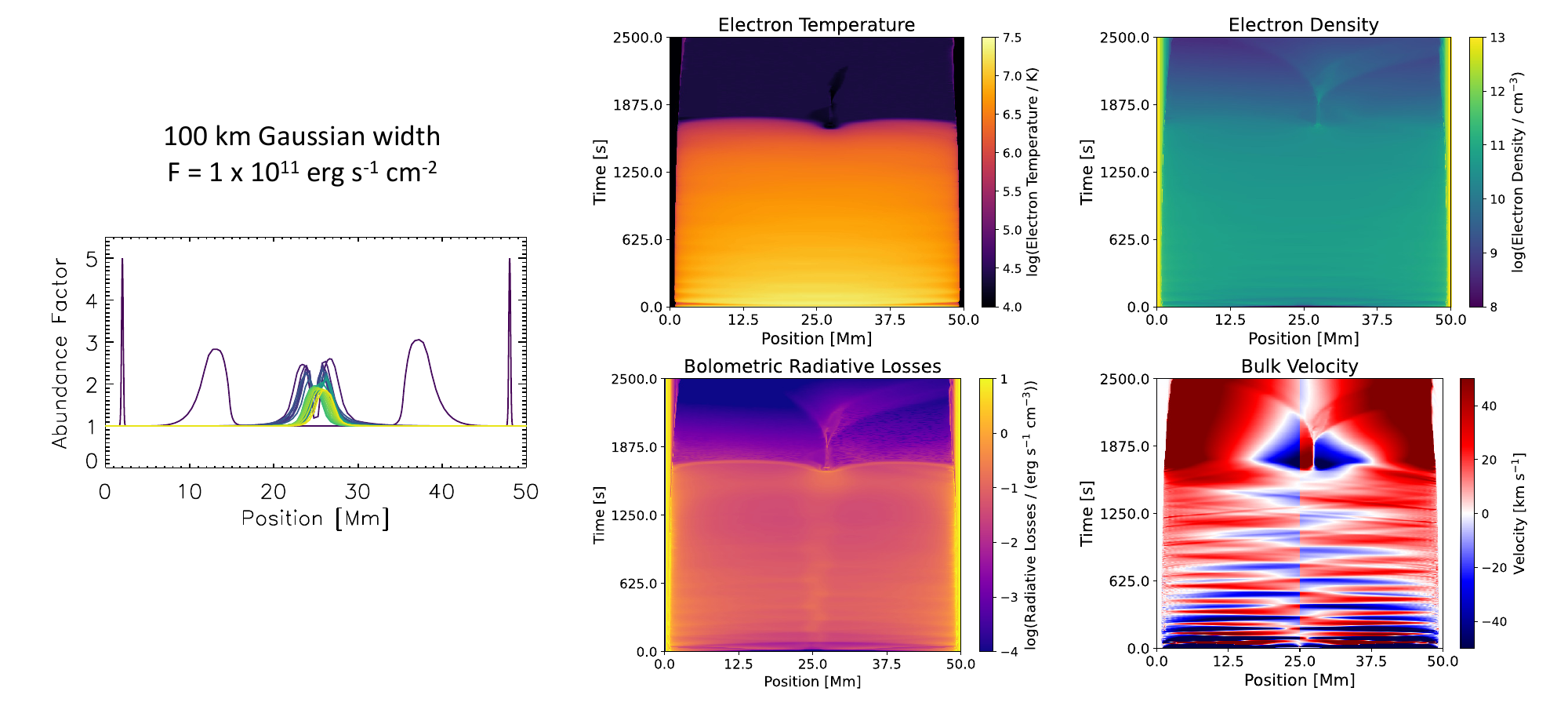}
    \includegraphics[width=\textwidth]{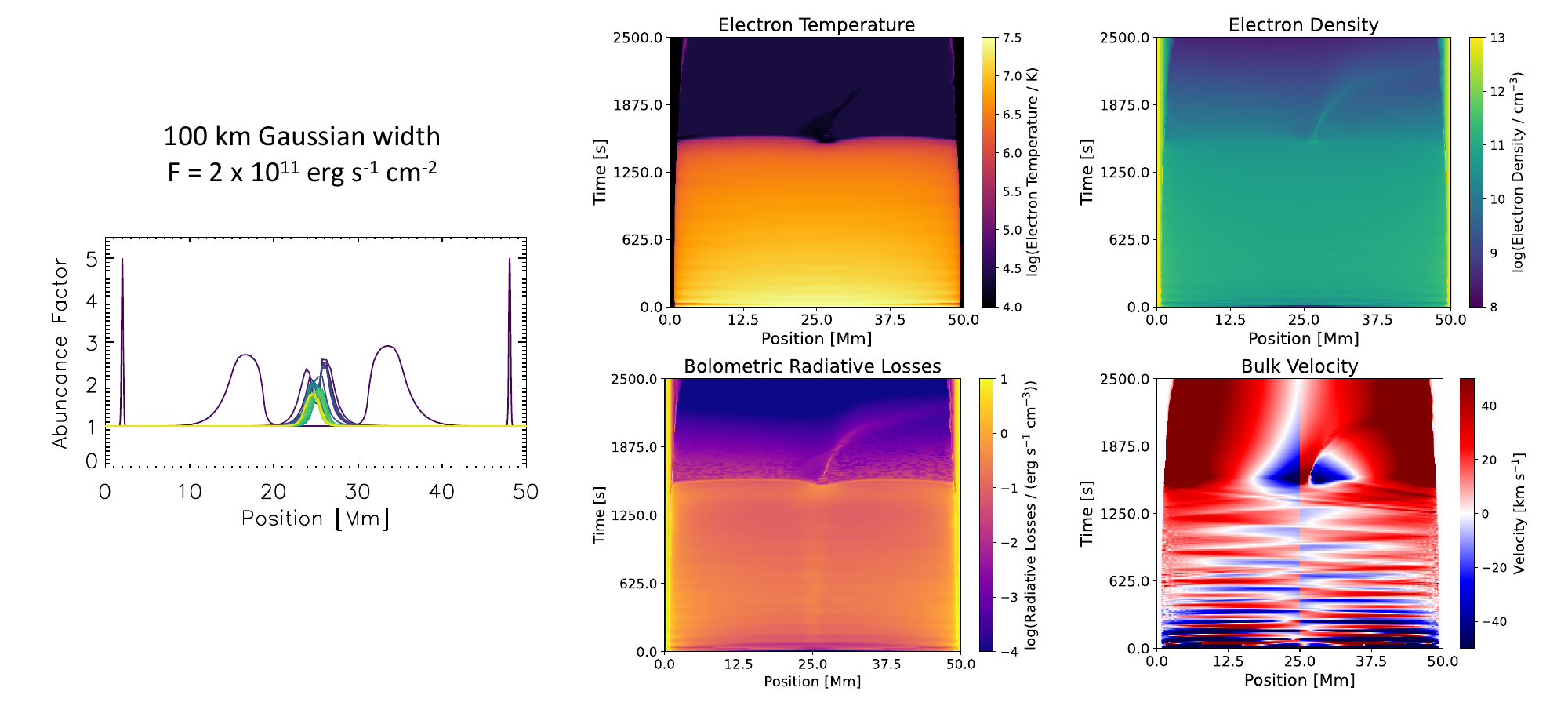}
    \caption{Continued for stronger heating cases.}
    \label{fig:narrow2}
\end{figure}

In these simulations, the evaporation additionally moves the fractionated material at the chromospheric footpoints into the corona, causing a localized variation in elemental abundance.  The flows carry that low-FIP enhanced plasma towards the apex, where it remains after evaporation ceases.  As a consequence, the majority of the coronal portion of the loop remains photospheric, except for the slight enhancement in $f$ near the apex of the loop.  That the peak occurs at the apex is a consequence of the assumptions of symmetry in both the initial distribution of $f$ and in the heating.  We examine asymmetries briefly further down, but significantly more work is required to understand how they impact the distribution of abundances.  The peak in $f \approx 2$ at the apex of the loop causes a slight enhancement in the radiative loss rate there, which can be seen as the local maximum in the radiative loss plot.  As the plasma cools, this directly induces a coronal rain event (similar to the simulations in \cite{benavitz2025}).  As the temperature falls locally, the radiative loss rate grows, causing the temperature to plummet ever faster.  The dip in temperature causes an inwards flow of material so that the density increases, strengthening the cooling further.  The rain can be seen in the plots corresponding to the locations with a local increase in density and radiative loss rate, and a dip in the temperature.  The formation of rain becomes noticeable around 1.3 MK in each case, with respective densities of $9 \times 10^{9}, 2 \times 10^{10}, 4 \times 10^{10}, 5 \times 10^{10}$ cm$^{-3}$.  The flows accelerate from around 20 to 50 km s$^{-1}$ as they precipitate.  Rain forms with all of the heating rates considered in this case, but the enhancement in $f$ becomes more localized with stronger evaporation.

The second case, with wide spikes, is shown in Figure \ref{fig:wide}.  The plots are similar to the previous figure.  In this case, because the initial enhancement in $f$ extends much deeper into the chromosphere, there is significantly more low-FIP enhanced plasma that can be transported into the corona.  This results in stronger radiation than in the previous case, which allows the loops to cool noticeably faster.  The early evolution of these loops occurs similarly to the previous case: before the density rises significantly, radiation is relatively weak compared to cooling by thermal conduction.  In the weakest heating case, the evaporation event carries enough low-FIP enhanced plasma into the corona that it becomes fully fractionated, with $f \approx 4$ across the entire coronal segment of the loop, only a minute after the onset of heating.  As a result, the radiative loss rate becomes enhanced almost uniformly across the corona, except for a small local dip near the apex.  The material in the corona prior to evaporation, with $f = 1$, is pushed towards the apex, causing this dip, essentially the opposite of the scenario presented in \cite{benavitz2025}.  Because the radiative cooling is nearly uniform, the loop cools as a whole, and no coronal rain forms (see also \cite{reep2020}).  With stronger heating rates, though, the low-FIP enhanced plasma becomes more and more compressed near the apex of the loop, so the radiative loss rate is not uniform across the corona, but peaked near the apex.  Coronal rain only forms in the strongest heating cases ($F_{0} \gtrsim 10^{11}$ erg s$^{-1}$ cm$^{-2}$), suggesting that the material must be sufficiently compressed for a localized runaway cooling to occur.  With a heating rate of $10^{11}$ erg s$^{-1}$ cm$^{-2}$, the rain becomes noticeable around 1.3 MK with a density of around $3 \times 10^{10}$ cm$^{-3}$, while in the strongest heating case, the rain becomes noticeable around 2 MK with a density of around $6 \times 10^{10}$ cm$^{-3}$.
\begin{figure}[!h]
\centering
\includegraphics[width=\textwidth]{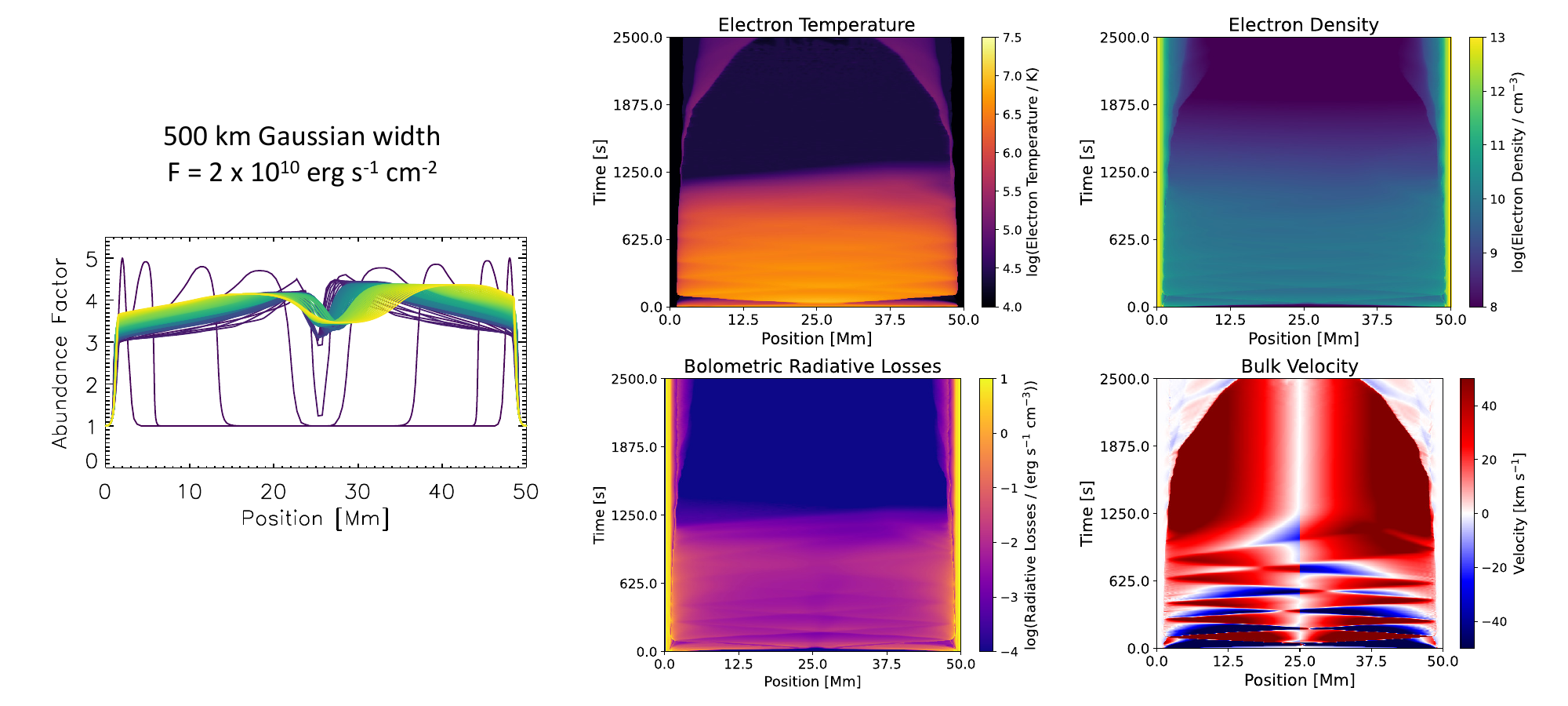}
\includegraphics[width=\textwidth]{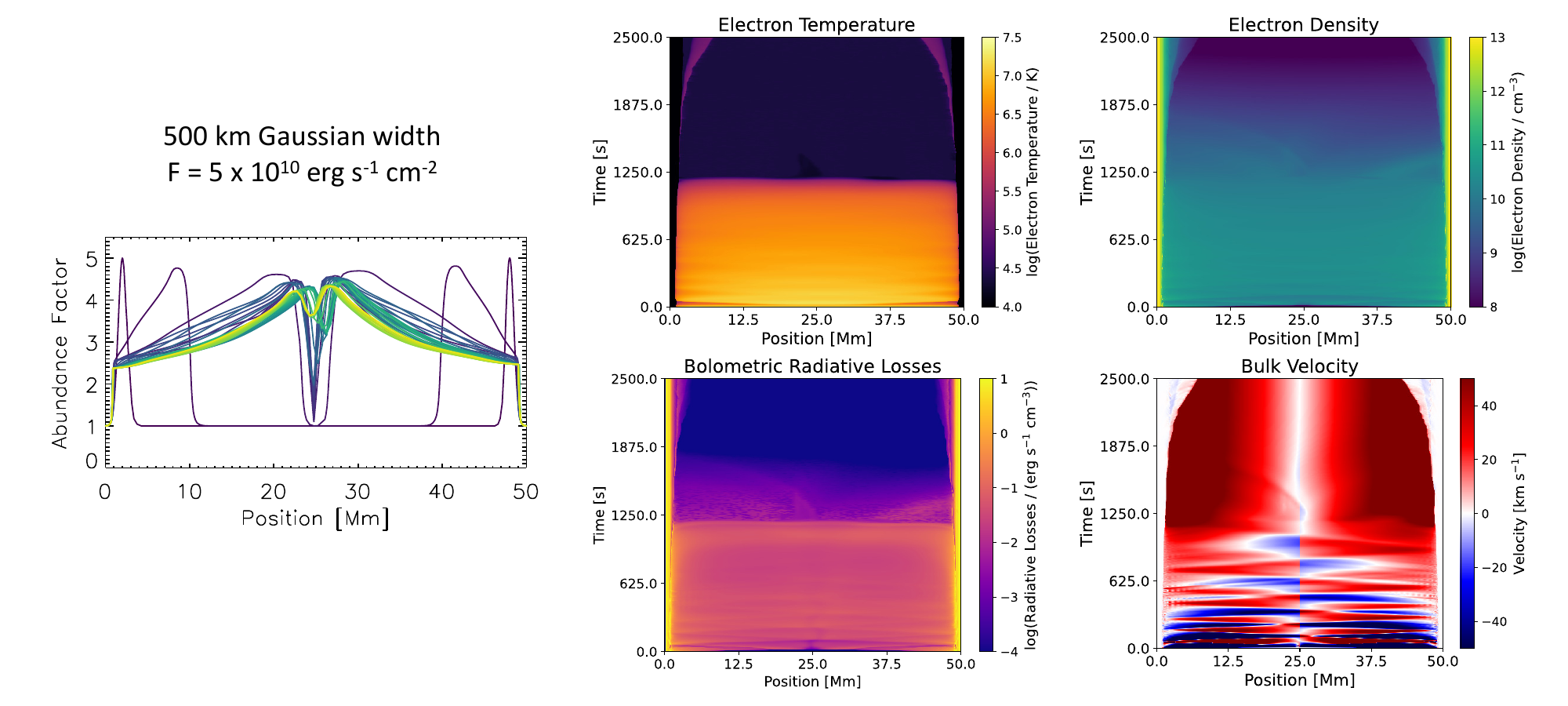}
\caption{Similar to Figure \ref{fig:narrow}, with wide spikes in the abundance factor $f$ at the footpoints of the loops.  In this case, since the enhancement contains a much deeper portion of the chromosphere, the evaporation event carries up significantly more low-FIP enhanced plasma.  With weak heating, this creates a fully fractionated corona.  As the heating rate grows stronger, the evaporative flows compress more material near the apex, causing a slight localized enhancement there.  Only with the strongest heating do we see the formation of coronal rain.  (Figure continued on next page.)}
\label{fig:wide}
\end{figure}
\begin{figure}[!h]
\ContinuedFloat
\centering
\includegraphics[width=\textwidth]{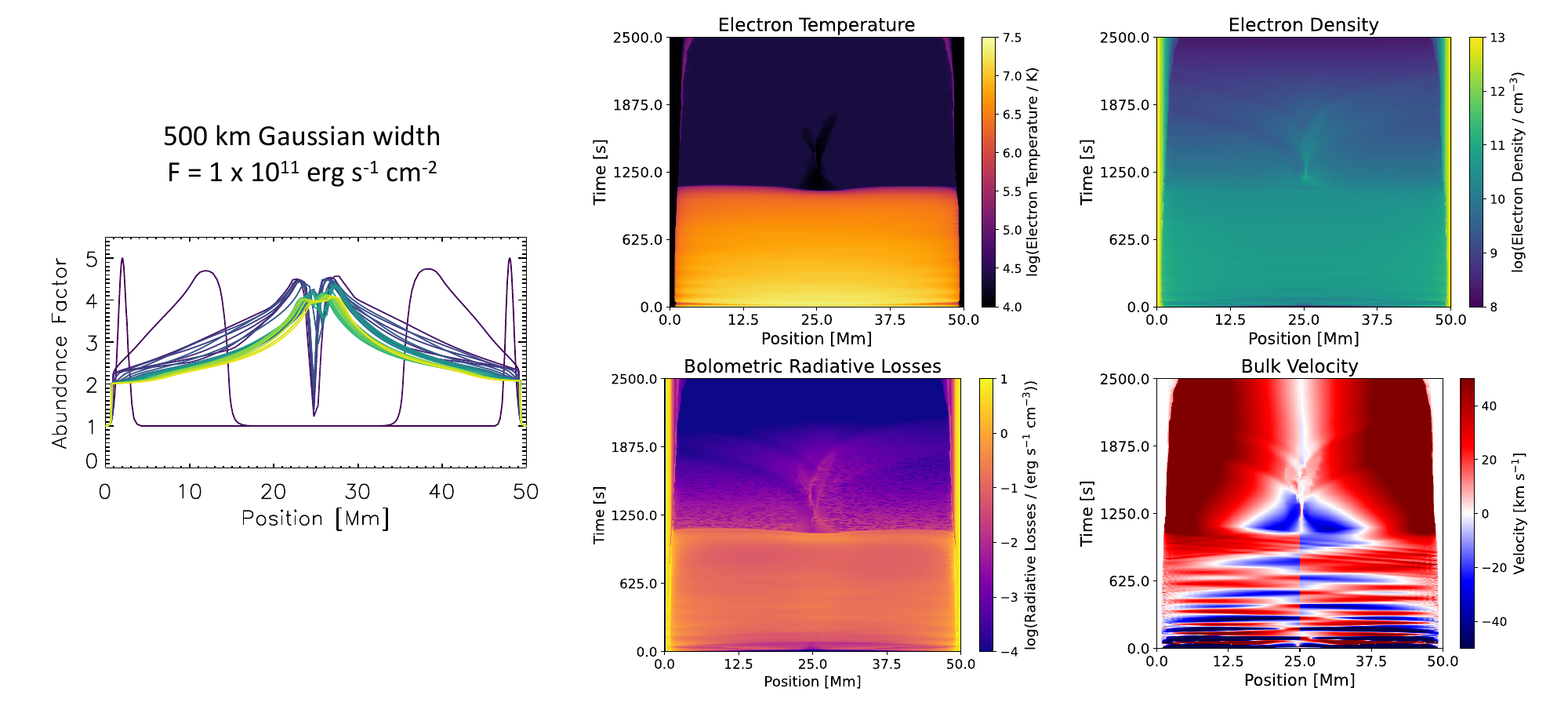}
\includegraphics[width=\textwidth]{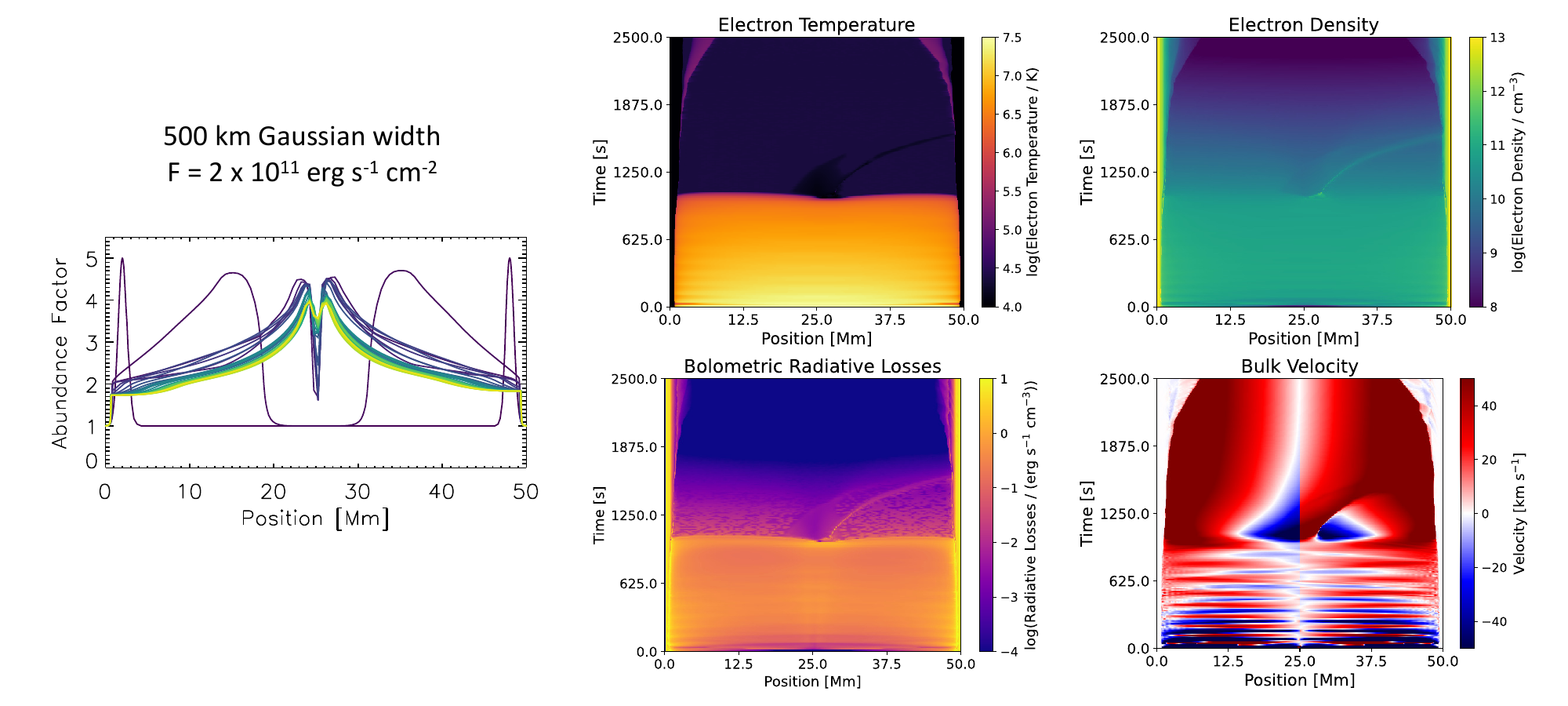}
\caption{Continued for stronger heating cases.}
\end{figure}

In both Figure \ref{fig:narrow} and \ref{fig:wide}, as well as in \cite{benavitz2025}, the heating rates and initial distributions of abundances were assumed to be symmetric, which is unlikely to be realistic.  We briefly examine the impact of asymmetries on these results.  We begin with the assumption of asymmetric heating: what happens if one leg receives more heat than the other?  In \cite{reep2020}, it was demonstrated that asymmetric heating cannot cause coronal rain when assuming fixed abundances, but the role of asymmetries has not been examined with spatially-varying abundances.

\begin{figure}[!h]
\centering
\includegraphics[width=\textwidth]{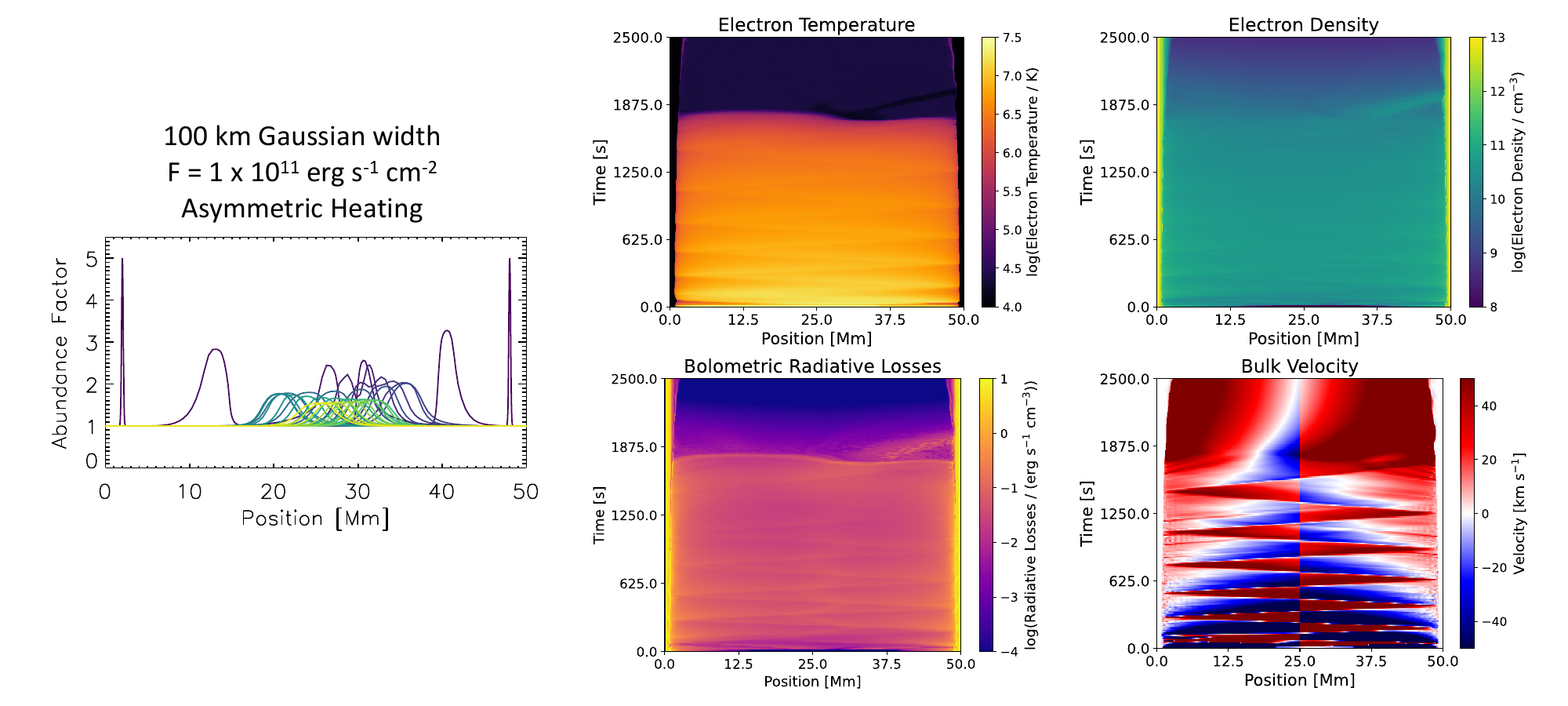}
\includegraphics[width=\textwidth]{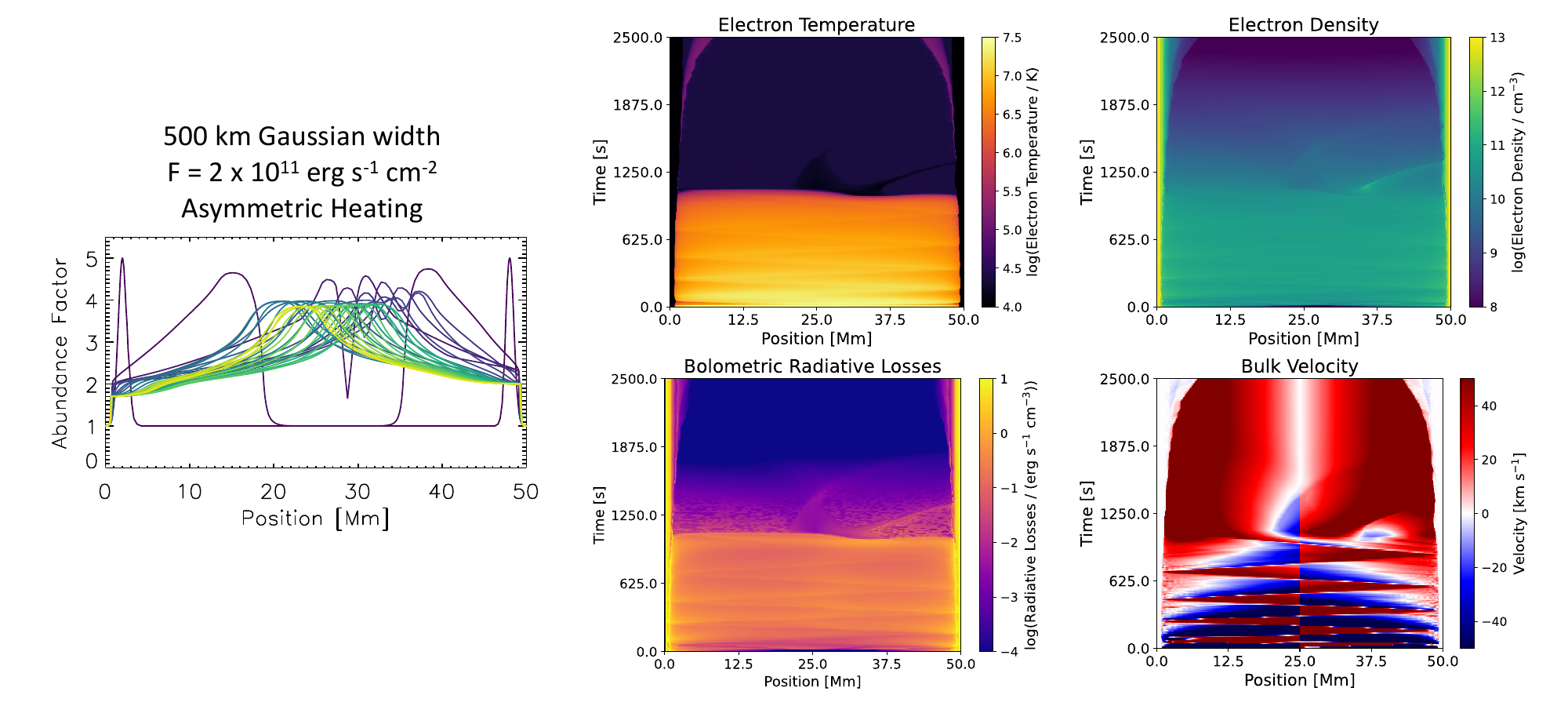}
\caption{Similar to Figure \ref{fig:narrow}, showing two examples with asymmetric heating, with the right-hand leg of the loop receiving half the energy as the left-hand leg.  The top simulation shows an example of two narrow spikes at the footpoints, while the bottom simulation shows a case with wide spikes at the footpoints.  The asymmetry in the heating affects whether rain forms where on the loop it forms.  The supplemental material includes plots and simulations with the other energy fluxes considered. }
\label{fig:asymm_heating}
\end{figure}

In Figure \ref{fig:asymm_heating}, we show two examples with asymmetric heating.  The top panels show a simulation with two narrow spikes at the footpoints, heated by a beam with energy $10^{11}$ erg s$^{-1}$ cm$^{-2}$ on the left hand leg of the loop, and half that amount on the right hand leg.  The evolution of temperature, density, and velocity are similar to the previous cases shown in Figure \ref{fig:narrow}.  The abundance factor, however, becomes peaked slightly to the right of the apex, such that the coronal rain forms on the right hand side of the loop, at around $s = 37$ Mm, when the loop cools to around 1 MK.  In the supplemental material, we also show the other energy fluxes, but in this case, rain does not form with $2 \times 10^{10}$ erg s$^{-1}$ cm$^{-2}$ because the abundance factor is not sufficiently concentrated to one location.   The bottom panels show an example with wide spikes at the footpoints.  The evolution of the basic variables is similar to Figure \ref{fig:wide}.  The abundance factor becomes strongly compressed due to the strong flows, which allows the radiation to peak locally and form rain, at around $s = 35$ Mm.  However, rain does not form in cases with weaker heating since the radiation is not sufficiently localized.  In general, we see that asymmetric heating shifts the location where coronal rain can form.  In some of the cases, additionally, we not see rain form at all because the colliding flows do not sufficiently compress the plasma.  

\begin{figure}[!h]
\centering
\includegraphics[width=\textwidth]{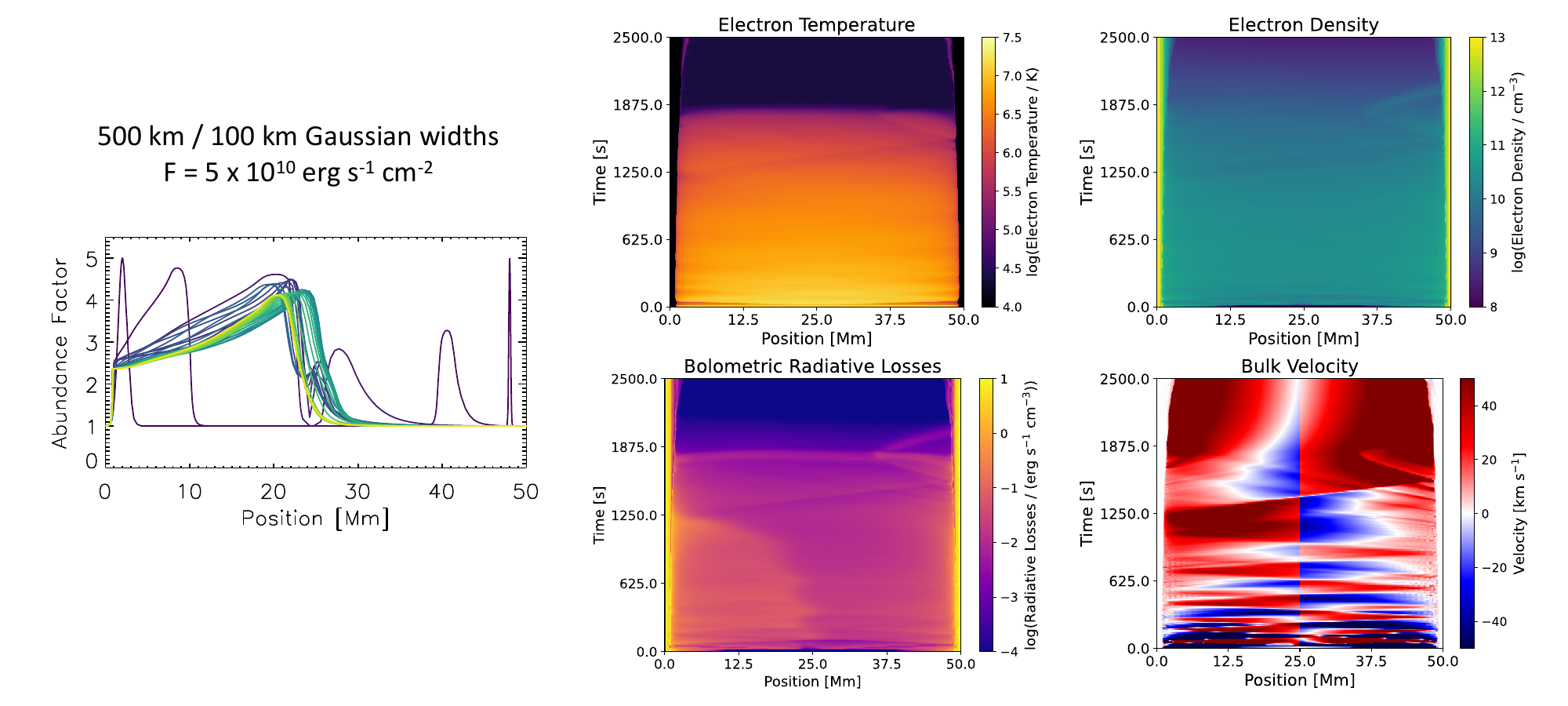}
\includegraphics[width=\textwidth]{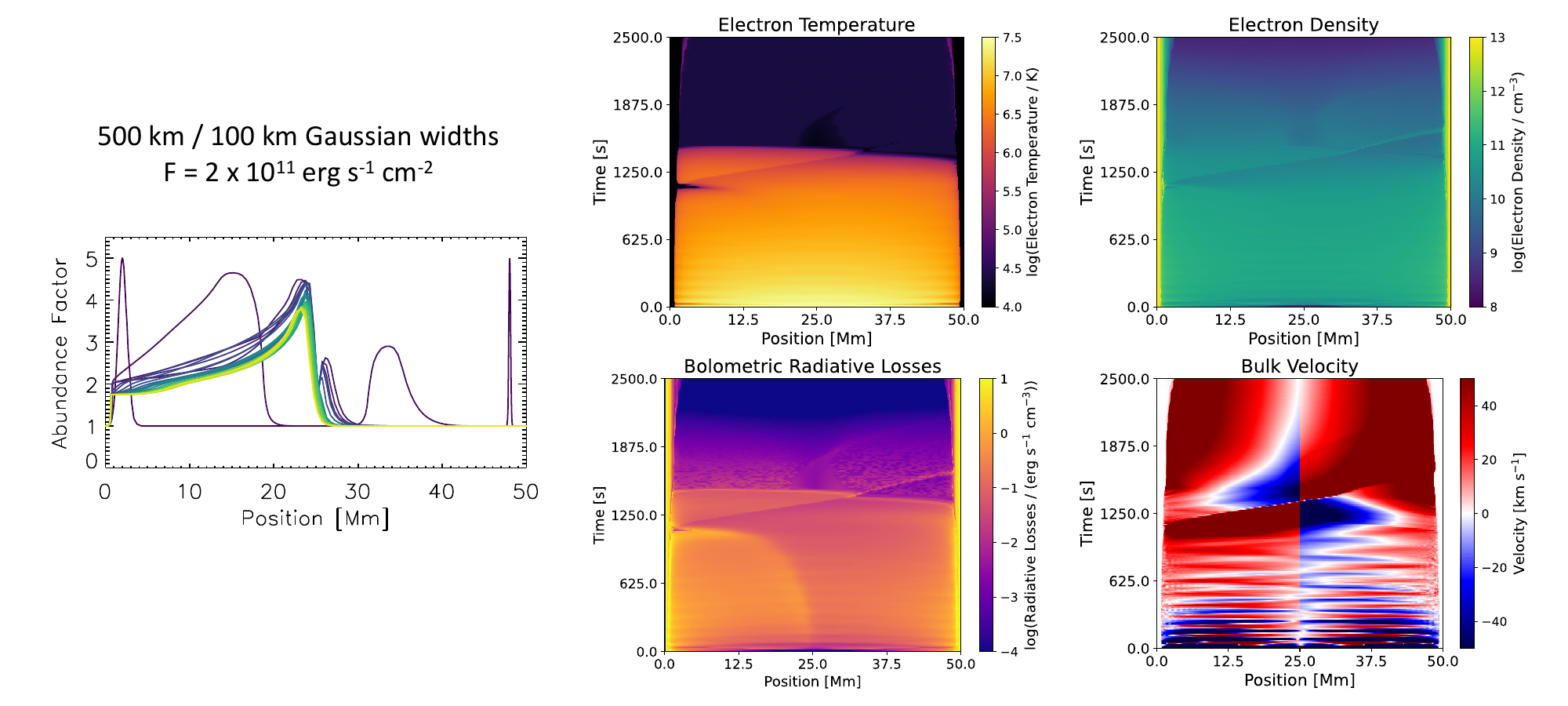}
\caption{Similar to Figure \ref{fig:narrow}, showing two examples with asymmetric footpoints, with the left leg having a 500 km width and the right leg having a 100 km width.  The top simulation shows a heating rate of $5 \times 10^{10}$ erg s$^{-1}$ cm$^{-2}$, while the bottom simulation shows $2 \times 10^{11}$ erg s$^{-1}$ cm$^{-2}$.  The abundance factor becomes peaked near the apex early on, but the low-FIP enhanced plasma later falls towards the left leg (readily seen in the radiative loss plots).  Rain forms with the stronger heating cases, but not with weaker heating, though the loop's evolution is similar.}
\label{fig:asymm_FP}
\end{figure}

Finally, we examine asymmetries in the initial distribution of abundance factor.  What happens if one footpoint is fractionated at different depths than the other?  Figure \ref{fig:asymm_FP} shows two examples of loops with asymmetric footpoints: the left footpoint has a wide spike while the right footpoint has a narrow spike.  The heating is assumed to be symmetric in this case.  The top plots show a weaker heating case, with energy flux $5 \times 10^{10}$ erg s$^{-1}$ cm$^{-2}$, while the bottom plots show a stronger heating case with $2 \times 10^{11}$ erg s$^{-1}$ cm$^{-2}$.  The evolution of the temperature, density, and velocity are similar.  Evaporation pushes the low-FIP enhanced plasma from the footpoints into the corona, where the two shock fronts collide at the apex.  The abundance factor becomes peaked near the apex as a result, but eventually pressure waves push that material towards the left-hand footpoint, which can be easily seen in the radiative loss plots.  With stronger heating, the density is higher, so this plasma cools sufficiently to trigger the catastrophic cooling that causes rain to form near the left-hand footpoint and quickly precipitate to the chromosphere, where the low-FIP enhanced plasma then remains.  The impact at the chromosphere also triggers a recoil of relatively high density and high temperature material that propagates across the loop to the other footpoint.  We conclude that the asymmetric distribution of low-FIP enhanced plasma affects the dynamics quite significantly, which requires further examination in the future.  

\section{Discussion}

The evolution of elemental abundances in the solar atmosphere is critical to understanding the radiative output of the corona.  Observations have shown that abundances vary both in space and time, but the mechanisms causing this remain poorly understood.  The FIP effect, an enhancement in the corona of elements with FIP less than 10 eV, is suspected to be caused by a ponderomotive acceleration induced by propagating Alfv\'enic waves \cite{laming2015,martinez2023}.  This acceleration, perhaps of order 1-10 km s$^{-2}$ \cite{laming2021}, causes a slow diffusion of low-FIP enhanced plasma into the corona, which would produce the FIP effect over relatively long time scales (hours to days).  Observations of quiescent active regions show similar fractionation timescales, ranging from hours \cite{mondal2023} to days \cite{baker2018}.  However, if there were fractionation at the footpoints, an evaporation event could carry that material into the corona much more rapidly, perhaps in mere minutes.

We have examined this idea with hydrodynamic simulations.  We do not simulate the FIP effect directly, but assume that there is an enhancement of low-FIP enhanced plasma at the footpoints due to a ponderomotive acceleration.  We then set off strong, impulsive heating events that cause a rapid flow into the corona.  The amount of low-FIP enhanced plasma transported into the corona depends on how deep the fractionation process occurs.  With relatively narrow spikes in fractionation, the coronal segment of the loop remains photospheric, with a slight enhancement in $f$ near the apex.  With wider spikes fractionating material deeper in the chromosphere, the loop can become fully coronal in abundance.  Asymmetries can affect whether rain forms, and, if so, where and when along the loops it may form, depending on how the low-FIP enhanced plasma is advected.  

Observations are necessary to verify or further constrain these models.  The distribution of the abundance factor with height in the solar atmosphere is a critical measurement.  For example, measurements of FIP bias $f$ with Ca/Ar and Fe/S ratios found an enhancement near the apex of post-flare loops \cite{to2024}.  However, those ratios are measured at temperatures around 2.5-3 MK, so they are only observed in the corona (and well above the temperature of coronal rain), but it is also necessary to determine how deep into the chromosphere fractionation can occur.  As the simulations here show, the resultant distribution of $f$ in the corona depends on this.  This was suggested as one possible explanation for differing abundance factors in post-flare rain and the loops in which it forms \cite{brooks2024}.  

There is thus an observational prediction: the presence or amount of coronal rain may be tied to both the fractionation depth and the strength of the heating.  This could be tested by examining coordinated observations of events with (1) good hard X-ray data to constrain the rate and depth of electron beam heating (\textit{e.g.} RHESSI \cite{lin2002}, SO/STIX \cite{krucker2020}, Fermi/LAT \cite{atwood2009}), (2) observations of coronal rain (\textit{e.g.} SDO/AIA 304 \cite{lemen2012}, IRIS/SJI \cite{depontieu2014}, DKIST/DL-NIRSP \cite{jaeggli2022}), and (3) spectral data to measure the abundance factor in space and time (\textit{e.g.} Hinode/EIS \cite{culhane2007}, MUSE \cite{depontieu2020}, EUVST \cite{shimizu2019}).  It would be illuminating to examine events of different flare classes to compare how the magnitude of heating affects the abundance factor evolution and the formation of rain.  It would also be important to contrast the observed properties of rain (densities, temperatures, morphology, etc.) with those predicted by the simulations, and refine the model-data comparison to better understand the process. 

\vskip6pt

\ack{All of the simulations used to produce the figures in this paper are available via Zenodo, \href{https://doi.org/10.5281/zenodo.17784826}{https://doi.org/10.5281/zenodo.17784826}.  We thank the Royal Society for its support throughout the organization of the Theo Murphy meeting and production of this special issue.  JML was supported by basic Research Funds of the Office of Naval Research.  We thank the referees for helpful insights that clarified and improved the text.}


\vskip2pc

\bibliographystyle{RS} 
\bibliography{references} 

\end{document}